\documentstyle[11pt,paspconf]{article}

\begin{document}
\title{BATSE observations of BL Lac Objects.}
\author{V. Connaughton}
\affil{NASA Marshall Space Flight Center, AL 35812}
\author{C.R. Robinson, M.L. McCollough}
\affil{Universities Space Research Association}
\author{S.A. Laurent-Muehleisen}
\affil{Lawrence Livermore National Laboratory}

\begin{abstract}
The Burst and Transient Source Experiment (BATSE) on the Compton
Gamma-Ray Observatory has been shown to be sensitive to non-transient
hard X-ray sources in our galaxy, down to flux levels of 100 mCrab for
daily measurements, 3 mCrab for integrations over several years.  We
use the continuous BATSE database and the Earth Occultation
technique to extract average flux values between 20 and 200 keV from
complete radio- and X-ray- selected BL Lac samples over a 2 year period.
\end{abstract}


\vspace{-0.3cm}
\section{Introduction}
\vspace{-0.3cm}

BL Lac objects have been studied throughout the electromagnetic spectrum;
from radio to TeV gamma rays
each energy band has provided clues to the energy production mechanisms
at the source.
The lower energy part of a BL Lac spectrum is thought to 
be synchrotron radiation of relativistic electrons in a jet 
oriented at a small angle to the observer,
and the higher energy emission Inverse Compton (IC)
scattering by the electrons either of these 
same synchrotron photons or photons external to the jet.

The position of the peak in the synchrotron emission is often used
to differentiate between two classes of BL Lac objects -  low-frequency
(LBL) and high-frequency (HBL) - the two
classes correspond roughly to those first
detected in radio (RBL), and those selected
in X-ray measurements (XBL) respectively.
The distinctions between the classes, however, are not always clear.  
Sambruna et al. (1996) find that a subclass
of LBLs look like HBLs viewed from a closer angle to the jet i.e. their
soft X-ray spectrum is dominated by synchrotron (like HBLs) but flat
enough to suggest an IC component (like LBLs). They   suggest that in
general
a flatter component becomes apparent in the spectra of LBLs above 1 keV which
may signify the start of the IC scattered component. The
corresponding IC spectrum of HBLs presumably
starts at tens of keV, but there appear
to be transition BL Lacs which exhibit properties of both classes.  
This
points toward there being a combination of beaming and intrinsic
source property factors affecting the emssion we see from these objects.

Studying BL Lac objects in the X-ray 
energy band  gives us a window into the synchrotron to
Compton turning point.
BATSE provides the opportunity to monitor many BL Lacs over a long
time line in hard X-rays.  This reduces the problems of poor sampling,
source flux variability and spectral variability.  
Malizia et al. (1998) looked the Piccinotti sample of AGN 
(which includes 4 HBLs) with BATSE over a three year period  and
showed sensitivity to at least some of these objects.  
We present here a preliminary analysis of two years of BATSE data 
for the X-ray  Einstein slew survey (Perlman et al. 1996),
and the complete 1 Jy radio-selected sample (Stickel et al. 1991).

\vspace{-0.3cm}
\section{BATSE observations.}
\vspace{-0.3cm}

The BATSE Large Area Detector (LAD)
data consist of counts in 16 energy channels between 10 keV and several
MeV.  
BATSE detects persistent emission from hard X-ray sources by looking
at changes in the continuous LAD background rates when the source
rises from and sets behind the Earth (Earth Occultation
Technique; Harmon et al. 1992).  In the $\sim 90$ minute CGRO
orbit, each source will contribute two steps, one each from a rise and
a setting.
The weakness of BL Lac sources requires the summing of many 
steps - in general one needs integration times of
several  months to a year to extract a signal, although flaring activity
can be seen on time-scales of days, at least in the case of Mrk 501
(Connaughton et al. 1998).
A BATSE survey of 34 supernova remnants (McCollough et al. 1992)
indicates that the sensitivity
of the technique over a long time line (nearly 5 years) 
is about 3 mCrab.

Although steps from known bright sources are measured and fit in the
Earth Occultation technique,  non-statistical fluctuations 
suggest careful cleaning of the data is necessary
before making any flux estimates.
Data were cleaned by
(i)checking for and flagging spacecraft pointing-dependent fluctuations due to
contaminating sources  in the limbs to the source and (ii)removing
outlier individual occultation steps to reduce systematic 
errors.    These outliers 
are usually attributable to pulsar activity in individual
detectors.  
Fifteen blank fields  which 
were randomly generated and selected only if they were devoid of known strong
X-ray sources were included in the occultation analysis.   
The proximity and number of bright, interfering sources in the blank
fields can be judged by the recurrence of interference patterns in the 
lightcurves as the orbit of the spacecraft precesses and the relative geometries
of source and interferer  recur.  It was seen that
sources closer than $5^\circ$ to the galactic plane, or within
a cone $45^\circ$ around the galactic center up to a latitude of 
$30^\circ$ are particularly vulnerable to contamination.
The 5  BL Lac sources and 4 blank fields which fall in this region 
are difficult to resolve from the numerous brighter galactic sources  
even with careful cleaning and have been removed from the analysis.
In the remaining blank fields which had no visible contamination, the
distribution of the significance of
flux estimates was symmetrical, but clearly non-Gaussian.  The errors on flux
measurements were increased by $65 \%$ of the value of the statistical
error to account for the presence of
systematic errors. 

Table 1 shows the
flux estimates over the first 2 years of the BATSE mission in the energy
range between 20 and 200 keV where the
background fitting in step calculations 
is most accurate and the response of the detectors
is efficient.  A photon index of -2.0 was used in the conversion of
counts to photons when folding through the detector response.
The flux errors are statistical
but the significances  are based on the total error.
A  $3 \sigma$ excess corresponds
to a flux level of $\sim 8 \times 10^{-11}$ erg cm$^{-2}$ s$^{-1}$.
No blank fields showed an excess at this significance level, while 9
of the BL Lacs were detected.  Of these 9 detections,
2 (1144-379 and 1147+245) are questionable because the
fluxes measured in the rising and setting limbs were not consistent at
a $2 \sigma_{stat}$ level. 

There is no correlation between BATSE flux and redshift, galactic
latitude, or soft X-ray flux.  If one characterizes each object
according to its degree of ``HBL-likeness", the ratio 
of the X-ray flux at 1 KeV to
the radio flux at 5 GHz, 
where $log(F_x/F_r) = -5.5$ is the common dividing line
between LBLs and HBLs, 
one finds that while
BATSE appears sensitive to both HBLs and LBLs, the flux detected in the
HBLs appears greater.   Clearly, one needs more detections to quantify
this in a meaningful fashion. 

{\footnotesize
\begin{table}
\caption{Average BATSE fluxes 20-200 keV  between 1991 and 1993}
\begin{tabular}{||l|c|c|l|c|c||}
Source & Flux  $\times 10^{-11}$ & $N_\sigma$ & Source & 
  Flux $\times 10^{-11}$ & $N_\sigma$ \\
    &   erg cm$^{-2}$ s $^{-1}$  
   & & 
    &    erg cm$^{-2}$ s $^{-1}$  
   &  \\
\hline
blx0507-040  &  -8.13  $\pm$    1.51  &  -3.26 &	blx1517plus656  &  3.25  $\pm$   1.79  &  1.1 \\
blank 13  &  -6.22  $\pm$    1.48  &  -2.54 &	blx0158plus003  &  2.53  $\pm$   1.37  &  1.11 \\
blx2343-151  &  -3.85  $\pm$    1.42  &  -1.64 &	blr0851plus203  &  2.67  $\pm$   1.38  &  1.17 \\
blx0715-259  &  -3.64  $\pm$    1.47  &  -1.5 &	blx2005-489  &  2.97  $\pm$   1.52  &  1.18 \\
blank 4  &  -3.44  $\pm$    1.44  &  -1.44 &	blx1727plus502  &  3.37  $\pm$   1.7  &  1.2 \\
blank 3  &  -4.05  $\pm$    1.88  &  -1.3 &	blx1248-296  &  3.11  $\pm$   1.56  &  1.2 \\
blank 5  &  -2.94  $\pm$    1.44  &  -1.23 &	blx0502plus675  &  3.69  $\pm$   1.79  &  1.24 \\
blr2131-021  &  -2.37  $\pm$    1.42  &  -1.01 &	blx1028plus511  &  3.52  $\pm$   1.61  &  1.32 \\
blank 11  &  -2.37  $\pm$    1.45  &  -0.99 &	blx1553plus113  &  3.47  $\pm$   1.58  &  1.33 \\
blr0454plus844  &  -2.14  $\pm$    1.75  &  -0.74 &	blr1308plus326  &  3.22  $\pm$   1.46  &  1.33 \\
blr0235plus164  &  -1.66  $\pm$    1.42  &  -0.7 &	blr0820plus225  &  3.18  $\pm$   1.4  &  1.37 \\
blr0048-097  &  -1.54  $\pm$    1.39  &  -0.67 &	blx1959plus650  &  4.47  $\pm$   1.9  &  1.42 \\
blank 12  &  -1.56  $\pm$    1.43  &  -0.66 &	blx1101plus384  &  3.4  $\pm$   1.44  &  1.43 \\
blr0138-097  &  -1.13  $\pm$    1.39  &  -0.49 &	blr0537-441  &  3.74  $\pm$   1.54  &  1.47 \\
blx0806plus524  &  -1.3  $\pm$    1.64  &  -0.48 &	blank 8  &  3.59  $\pm$   1.43  &  1.52 \\
blx1215plus303  &  -1.04  $\pm$    1.59  &  -0.39 &	blr1418plus546  &  4.33  $\pm$   1.68  &  1.56 \\
blx1239plus069  &  -0.78  $\pm$    1.45  &  -0.32 &	blr1807plus698  &  5.41  $\pm$   1.95  &  1.68 \\
blx0548-322  &  -0.771  $\pm$    1.46  &  -0.32 &	blr2254plus074  &  4.01  $\pm$   1.37  &  1.77 \\
blx0347-121  &  -0.592  $\pm$    1.48  &  -0.24 &	blx0323plus022  &  4.15  $\pm$   1.42  &  1.77 \\
blx1312-423  &  -0.61  $\pm$    1.66  &  -0.22 &	blx1853plus671  &  5.55  $\pm$   1.88  &  1.78 \\
blx1544plus820  &  -0.259  $\pm$    1.88  &  -0.08 &	blr0823plus033  &  4.42  $\pm$   1.43  &  1.87 \\
blr1823plus568  &  -0.152  $\pm$    1.78  &  -0.05 &	blx1332-295  &  5.31  $\pm$   1.6  &  2.01 \\
blx0145plus138  &  0.0719  $\pm$    1.39  &  0.03 &	blx0219plus428  &  5.32  $\pm$   1.57  &  2.05 \\
blx0414plus009  &  0.299  $\pm$    1.49  &  0.12 &	blank 14  &  5.83  $\pm$   1.67  &  2.11 \\
blx0525plus713  &  0.431  $\pm$    1.84  &  0.14 &	blx2321plus419  &  5.73  $\pm$   1.54  &  2.25 \\
blx1218plus285  &  0.663  $\pm$    1.57  &  0.25 &	blx0229plus200  &  5.33  $\pm$   1.42  &  2.27 \\
blr2240-260  &  0.616  $\pm$    1.43  &  0.26 &	blr2007plus777  &  7.57  $\pm$   1.99  &  2.3 \\
blx0950plus495  &  0.678  $\pm$    1.57  &  0.26 &	blx1440plus122  &  5.51  $\pm$   1.43  &  2.33 \\
blr0716plus714  &  1.01  $\pm$    1.8  &  0.34 &	blx2155-304  &  5.41  $\pm$   1.4  &  2.34 \\
blx2326plus174  &  0.864  $\pm$    1.37  &  0.38 &	blx0927plus500  &  6.15  $\pm$   1.57  &  2.37 \\
blank 6  &  1.17  $\pm$    1.65  &  0.42 &	blr2200plus420  &  6.23  $\pm$   1.55  &  2.43 \\
blr0814plus425  &  1.09  $\pm$    1.5  &  0.44 &	blx2344plus514  &  6.8  $\pm$   1.67  &  2.46 \\
blr0735plus178  &  1.23  $\pm$    1.43  &  0.52 &	blx0647plus250  &  6.15  $\pm$   1.47  &  2.53 \\
blr0828plus493  &  1.4  $\pm$    1.58  &  0.53 &	blr0954plus658  &  7.45  $\pm$   1.74  &  2.59 \\
blank 10  &  1.56  $\pm$    1.7  &  0.55 &	blr1514-241  &  6.4  $\pm$   1.49  &  2.6 \\
blx1101-232  &  1.49  $\pm$    1.5  &  0.6 &	blx0737plus746  &  8.06  $\pm$   1.86  &  2.62 \\
blank 2  &  1.58  $\pm$    1.56  &  0.61 &	blr0118-272  &  6.59  $\pm$   1.41  &  2.83 \\
blx1118plus424  &  1.57  $\pm$    1.54  &  0.61 &	blr1147plus245  &  7.6  $\pm$   1.51  &  3.05 \\
blx1320plus084  &  1.49  $\pm$    1.42  &  0.63 &	blr1803plus784  &  10.4  $\pm$   2.01  &  3.13 \\
blx1218plus304  &  1.76  $\pm$    1.6  &  0.66 &	blr1538plus149  &  8.26  $\pm$   1.57  &  3.18 \\
blx1106plus244  &  1.73  $\pm$    1.49  &  0.7 &	blr1144-379  &  8.09  $\pm$   1.52  &  3.22 \\
blr0426-380  &  1.78  $\pm$    1.47  &  0.73 &	blx1652plus398  &  9.28  $\pm$   1.59  &  3.53 \\
blx1011plus496  &  1.97  $\pm$    1.58  &  0.75 &	blx1255plus244  &  9.03  $\pm$   1.47  &  3.72 \\
blx1133plus704  &  2.54  $\pm$    1.81  &  0.85 &	blx1212plus078  &  9.88  $\pm$   1.48  &  4.04 \\
blx1402plus042  &  2.08  $\pm$    1.4  &  0.9 &	blx1426plus428  &  10.9  $\pm$   1.54  &  4.28 \\
blx1421plus582  &  2.71  $\pm$    1.75  &  0.93 &	blx0120plus340  &  12.5  $\pm$   1.46  &  5.18 \\
blx1533plus535  &  2.95  $\pm$    1.68  &  1.06 & & & \\	
\end{tabular}
\end{table}}

The analysis outlined here is being applied to more samples of BL Lacs and to
the remaining 5 years of BATSE archival data.    
Detection of  LBLs, HBLs  and the intermediate BL Lacs
(Laurent-Muehleisen et al. 1997)
in significant numbers  may allow us to
to distinguish (by fitting spectra to the count rates rather
than assuming a -2.0 power law) between the hypotheses of 
(i)different IC components for distinct populations (Ghisellini 1999)
 and (ii)BL Lacs forming one family showing a continuum 
of properties.
  
\vspace{-0.3cm}
\section{Variability}
\vspace{-0.3cm}
The detection by BATSE of a hard X-ray Mrk 501 flare with a flat
spectrum out to at least 100 keV opens
the possibility that BATSE will be able to detect some sources in high
intensity states over time-scales of days.
If these flares are discovered as they are happening, BATSE might be used as
a monitor for strong, hard flares - multi-wavelength campaigns could
be initiated providing the detection is convincing.
A preliminary search of the 1991-93 database for day-scale flares (for
all 87 sources) 
of magnitude equal or greater than that seen in Mrk 501 in April 1997
yields only one convincing detection:
again, the object is Mrk 501.  It is clear that a more systematic search for
longer time-scale flaring may be fruitful.  One can see in
Table 1 that while the blank field excess distributions are 
symmetrical, there are far more excesses than 
deficits in the BL Lac samples, and many more $2\sigma$ excesses 
than expected by chance.  Since these sources have exhibited variability over
most time-scales at
all wavelengths at which they have been monitored, it is
 expected that those
sources which may not prove significant over integrations of 2 years might
be active on shorter time-scales.

\end{document}